# Study on the numerical value and the distance of symptoms on the Medical Decision Support System

**Won-Il Song, Taek-Jong Kim**

*Cutting-edge Science Institute, **Kim Il Sung** University,
Pyongyang, DPR Korea*

**Abstract**

It is an important subject how deal with the symptom's data, input data, to improve the accuracy and efficiency of the diagnostic algorithm in the medical decision support systems. In this paper, we described a method for the numerical value of symptoms to develop the medical decision support system that diagnose a large number of diseases, not one or two diseases. The numerical value of symptoms is realized by dividing the symptom into several parts in consideration of the following contents and setting up the specific value in each part. That is, the symptom is appeared in where, when the symptom is appeared, what about the symptom, How long the symptom is, etc. Then we decided the distance(similarity) between symptoms on the basis of the numerical value of symptoms. Determination of the distance between symptoms by numerical value enables estimating the symptoms of which disease is nearer from the symptoms of patient and obtaining the suitable diagnostic result.
So, we established useful method of high accuracy to diagnose various diseases revealed by different symptoms.

## 1. Introduction

Medical diagnosis of diseases is one of the most foremost important issues in the healthcare unit[1]. Usually the medical diagnosis is enable to make a wrong diagnosis and take much time and very expensive because the relation between diseases and symptoms is very complex. So, the research to diagnose the diseases using computer aided techniques progress actively in medical field[1-14]. Using computer aided techniques in medical field could reduce the medical error, time and cost[2]. The medical decision support system(MDSS) using computer aided techniques is designed to assist physicians and other health professionals with decision making tasks, such as determining diagnosis of patient's symptom data[15]. But the amounts of symptom data, the input data of MDSS, are so complex and voluminous that it is necessary to transform these mounds of symptom data into useful information for diagnosis[15]. Many data mining methodologies and technologies for numerical value of the symptom data are used in MDSS[1-3,15-25].

Manjusha et al. (2016) realized the numerical value of symptoms to predict the toxic epidermal necrolysis as given in Table 1[2].

**Table1.** Numerical value of symptoms for prediction of toxic epidermal necrolysis

| | Input Attributes or Symptoms for Prediction | |
|---|---|---|
| 1 | Age | (value 0:<40; value 1:>40) |
| 2 | Anticonvulsant therapy | (value 0:Yes; value 1:No) |
| 3 | Eye symptoms | (value 0:redness of eye; value 1:Conjunctivitis; value 3:Corneal involvement) |
| 4 | Blood pressure | (value 0:>120; value1:120-100; value 2:<100) |
| 5 | Temperature | (value 0:≤100; value 1:100-102; value 2:>102) |
| 6 | Nikolsky sign | (value 0:positive; value 1:Negative) |
| 7 | Pulse Rate | (value 0:<90; value 1:90-120; value 2:>120) |
| 8 | Respiratory Rate | (value 0:<15; value 1:15-20; value 2:>20) |
| 9 | Urea | (value 0:20-40; value 1:40-50; value 2:>50) |
| 10 | Bicarbonate level | (value 0: upto 20; Value 1:20-30; value 2:>30) |
| 11 | Leukocytosis | (value 0:<10 000; value 1:10 000-14 000; value 2:>14 000 |
| 12 | Area of involvement | (value 0:0-30; value 1:30-40; value 2:40-60; value 3:>60) |
| 13 | Skin Tenderness | (value 0:Mild; value 1:moderate; value 2:severe) |
| 14 | SGPT | (value 0:40-100; value 1:100-150; value 2:>150) |
| 15 | ALP | (value 0:140-150; value 1:150-200; value 3:>200) |

The 15 symptom indices related to the toxic epidermal necrolysis are assigned with 0~3 integral numbers by the base values. Then the severity prediction of toxic epidermal necrolysis is decided as four cases(No, Mild, Moderate, Severe) on the basis of numerical value like as above.

Navneet Walia et al. (2015) realize the numerical value of nine symptom indices for tuberculosis diagnosability as shown Table 2[1].

**Table 2.** Numerical value of symptoms for diagnosis of tuberculosis

| Input attribute | Data type | Acceptable score |
|---|---|---|
| Coughing | Integer | 0 mean < two weeks, 1 mean between two-three weeks, 2 mean > three weeks |
| BCG | Boolean | Yes = 0, No = 1 |
| Chest Pain | Boolean | No = 0, Yes = 1 |
| Malaises | Boolean | No = 0, Yes = 1 |
| Fever | Integer | 0 mean normal, 1 mean high, 2 mean subfebrile |
| Loss of appetite | Boolean | Yes = 0, No = 1 |
| Smoke addiction | Integer | 0 mean none, 1 mean less than eight, 2 mean eight to ten |
| Weight loss | Boolean | No = 0, Yes = 1 |
| Haemoptysis | Boolean | No = 0, Yes = 1 |

In the first group, the cough is categorized into three classes the patient have, '0' indicates a cough is less than two week, '1' indicates a cough is between two to three weeks, '2' indicates a cough is more than three weeks. BCG vaccine attribute shows that whether the patient has taken bacillus Calmette-Guerin vaccination or not. Chest pain, Malaises, loss of appetite and loss in weight has binary values. All these parameter has two values, either positive or negative. Smoke addiction parameter indicates a number of cigarettes consumed by a person per day. It consists of three subgroups, '0' indicates patient is a non-smoker, '1' indicates patient takes less than eight cigars per day, '2' indicates patient takes 6 to 10 cigars per day. Fever is classified into three classes, '0' means normal fever value which is nearly 36.5℃, '1' means fever value high, '2' means sub febrile fever value which exceeded 38.5℃. Haemoptysis parameter indicates there is coming of blood from respiratory tract of patient while coughing or not.

Sergei Parshutin et al. (2013) realize the numerical value of ten symptom indices for atrophic gastritis using Information Gain, GainRatio and Genetic Search[4].

Making the numerical values of symptoms as mentioned above is the method about some characteristic symptoms on a disease. Therefore, these methods can be used effectively to diagnose a disease but cannot be adopted on the diagnostic systems to diagnose many diseases collectively because a small number of the amounts of symptoms. And the practical patient's symptoms are very vague or complex but the symptoms considered above as symptoms peculiar to a relevant disease are so simple and plain that it's absurd on general diagnostic support systems. What is worse, there was nothing of special mention about the distance between symptoms of evaluating the similarity between symptoms.

We adopted here numerology of the symptoms and the distance between symptoms to develop the MDSS which can be taken synthetically in the diagnosis of many diseases where there are lots of numbers of symptoms and the symptoms are more concrete and complex.

## 2. Methods

### 2.1. Making the numerical value of symptoms

The numerical value of symptoms, the input data of MDSS, is the numerical expression of the contents. The symptom data of MDSS which diagnose lots of diseases in an all-round way is more concrete and complex than the symptom data of MDSS which diagnose a disease specially.

We divided the symptom data into several elements and considered the contents before making the numerical value. The symptoms consist of the elements like when, where, what is trouble, how long, in which direction, etc.

Namely, a symptom x can be formalized as follows:

$$x = \{c_1, c_2, \ldots, c_p\} \qquad c_i: \text{element}, \quad p: \text{number of elements}$$

Making the numerical value of symptoms is advanced by determining the element's value and calculating the symptom's characteristic value.

#### 2.1.1. Determine the element's value

The element's value $ch_i (i = \overline{1,p})$ of the $c_i$, the $i$th element of the symptom, is marked uniquely with a number of 1~4 figures whole according to the numbers of contents in the element.

It is as follows in the concrete.

(1) The 'where' element - Here, we divided a human body into several parts and assigned a different number of one figure to each. Next, we divided each parts into several small parts and assigned a different number of one figure to each. And we divided each small parts into several mini parts and assigned a different number of one figure to each, too. That is, we subdivided the human body into three steps and the division number is not over ten in each step. If the number of division is over ten, we can assign the number of two figures. The subdivision of the element 'where' shows in Figure 1. For example, the element's value of the 'head' is '100' and the element's value of the 'eye' is '120' and 'iris' is '123'.

(2) The elements of 'when', 'how serious', 'how long' etc. is assigned 0~9 number using fuzzy method and numerical method.

(3) The elements of 'what is trouble' and 'which direction move' is assigned 0~99 numerical value uniquely amenably to the degrees.

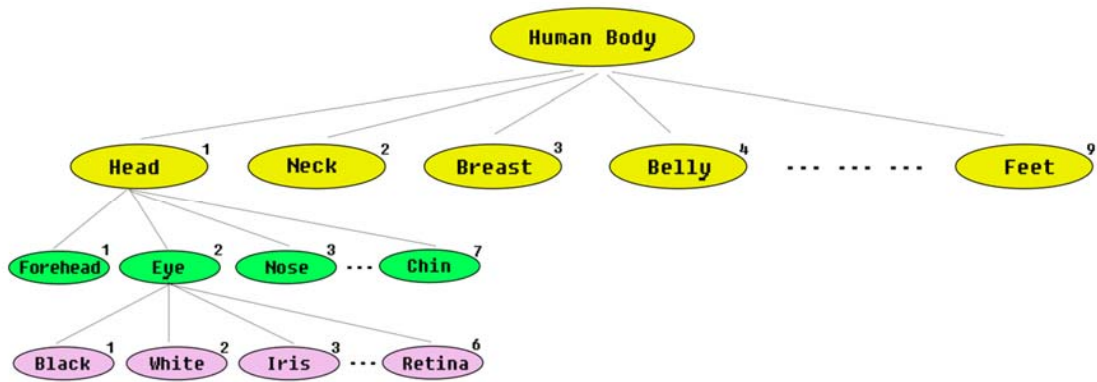

**Fig.1.** The subdivision of the element 'where' of a human body

The decision of element values for every symptoms was performed under the assistance of many physicians and health experts. When the cipher of $c_i$, the $i$th element of a symptom, is marked with $t_i$, $i = \overline{1,p}$, the numbers of the whole ciphers about the certain disease can be marked with $\sum_{i=1}^{p} t_i$.

**2.1.2. Calculate of the symptom's characteristic value**

The $ch(x_k)$, the characteristic value of a symptom $x_k$, is the value that combine every element's values.

It can be

$$ch(x_k) = \sum_{i=1}^{p} ch_i^{(k)} \times 10^{r_i}, k = \overline{1,q} \qquad (1)$$

Here, $r_i = \sum_{j=i+1}^{p} t_j$ and $q$ is whole number of the symptoms treated. For example, let's make the numerical value of a symptom $x_k$, "serious headache lasts for a long time.". This symptom consists of 4 elements that are "where"(head), "what is the trouble"(pain), "how"(serious) and "how long"(long time). We described that every element's value have determined uniquely previously.

If the $ch_1^{(k)}$ (the element's value of the "head")=100, $ch_2^{(k)}$ (the element's value of the "pain")=002, $ch_3^{(k)}$ (the element's value of the "serious")=3 and $ch_4^{(k)}$ (the element's value of the "long time")=4, the characteristic value of this symptom $ch(x_k) = 10000234$.

As the above, any symptom can be expressed by numerical value.

## 2.2 Setting the distance between symptoms

The distance between the symptoms can be calculated by using the characteristic values mentioned above. So, calculating the characteristic values of the symptoms can be regarded as a prerequisite for the calculating the distance between the symptoms. Let's see how to calculate the distance between the symptom $x_i$ and symptom $x_j$.

### 2.2.1. Distance between elements

First, calculate the distance between elements of the symptom. The distance between two elements is assigned a fixed positive number if these elements are not related to each other and if these elements are related, the distance is assigned a positive number within the limits according to the relation degree. For example, there is not related between the part 'belly' and the part 'leg' and there is related between the part 'leg' and the part 'feet'.

**Definition 1.** Distance between element $c_i^{(k)}(k = \overline{1,p})$ of symptom $x_i$ and element $c_j^{(k)}$ of symptom $x_j$ is defined as

$$d_k(x_i, x_j) = \begin{cases} d_{max_k}(d_{max_k} > 0), & i \neq j & \text{When have no relation} \\ 0 & i = j \\ d_{ij}(d_{ij} > 0), & i \neq j, d_{min_k} \leq d_{ij} \leq d_{max_k} & \text{When have relation} \end{cases} \quad (2)$$

Here, $d_{max_k}$ is the value that defines the upper limits of the distance between the $k$th elements of two symptoms and $d_{min_k}$ is the value that defines the lower limits of the distance between the $k$th elements of two symptoms.
And $d_{min_k} > d_{max_{k-1}} (k=\overline{2,p})$.

### 2.2.2. Distance between symptoms

After calculate the distance between the elements, the distance between two symptoms $x_i, x_j$ can be defined as follws.

**Definition 2.** About two symptoms $x_i, x_j \in X$

$$d(x_i, x_j) = \sqrt{\sum_{k=1}^{p}(d_k(x_i, x_j))^2} \tag{3}$$

is the distance between two symptoms $x_i, x_j \in X$.

In the definition 1 and definition 2, It is true that the distances between the elements and the symptoms satisfy an axiom of the distance.

## 3. Result

We have performed the numerical values of symptoms about 6000 symptoms of more than 1500 diseases such as internal diseases, surgery diseases, ophthalmic diseases, infectious diseases, etc as described in section 2.1.

The numerical value of every symptoms was performed under the assistance of many physicians and health experts as mentioned in section 2.1.1.

Some symptoms and their corresponding characteristic values are shown in table 3.

**Table 3.** Some symptoms and their corresponding characteristic values

| No | symptom | characteristic value |
|---|---|---|
| 1 | A mordant headache continues long in existence | 10000234 |
| 2 | A deglutition trouble grows violent in a brief period | 20000500 |
| 3 | A chest pain comes violently all of a sudden | 30001101 |
| 4 | My hands and feet are benumbed severely | 60040302 |
| 5 | A digestion trouble comes on a full stomach | 50030400 |
| ... | ... | ... |

The diagnosis algorithm can be maked as follows using the numerical value of symptoms and the distance between symptoms.

   Step 1. Compose the symptom list of each disease on the basis of making numerical value of symptoms.

   Step 2. Compose the symptom list of the patient's data in the same way.

   Step 3. Calculate every distance between the symptom list of patient and the symptom lists of diseases in database on the basis of distance between symptoms.(The details toward this will be discussed in the future.)

   Step 4. Decide the disease that has the shortest distance as the prediction of diagnosis.

The simple principle to diagnose a disease using the distance between symptoms is shown as Figure 2.

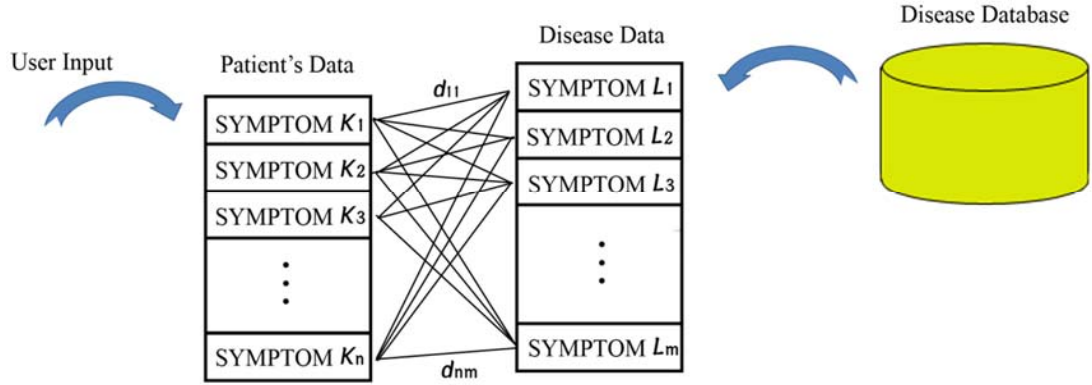

**Fig. 2.** The simple principle to diagnose a disease using the distance between symptoms

**4. Conclusion**

In this paper, by making the numerical value of symptoms, input data, and adopting the distance between every symptom in the medical decision support systems, analogy degree of the symptoms can be estimated in a quantitative way and this formed the foundation of evaluating the similarity between symptoms further. So, It can be possible to make the algorithm that diagnoses diseases based on the distance(similarity) between symptoms of diseases. Therefore we have established the correct methodology to diagnose diseases of the general areas synthetically as compared with previous one that diagnose a disease specially. The concrete description about the distance between symptom lists and the diagnostic algorithm will be discussed in next paper.